# Beyond Walker Breakdown through the Resonant Dissipation: Dramatic Enhancement of Magnetic Domain Wall Velocity via Resonant Excitation of Standing Wave Modes of Domain Wall Structure

Ganghwi Kim[1,*], Dae-Han Jung[1], Hee-Sung Han[2] and Ki-Suk Lee[1,†]

[1]*School of Materials Science and Engineering, Ulsan National Institute of Science Technology, Ulsan 44919, Republic of Korea*

[2]*Center for X-ray Optics, Lawrence Berkeley National Laboratory, Berkeley, CA 94720, USA.*

## Abstract

The dynamic behaviors of magnetic domain walls have significant implications for developing advanced spintronic devices. In this study, we investigate the intriguing resonance phenomenon within the magnetic domain wall structure and its profound influence on dynamic motion, focusing on the dissipation mechanism. By applying a static external magnetic field, we observe a remarkable amplification of domain wall velocity, surpassing the limitations of the conventional one-dimensional model. To quantify this enhancement, we introduce a novel parameter, the distortion variation rate, which captures the rapid and pronounced changes occurring within the domain wall structure. Through comprehensive micromagnetic simulations, we establish a robust relationship between speed and distortion variation rate, thereby validating our theoretical framework. Our findings provide crucial insights into the underlying mechanisms governing domain wall dynamics while paving the way for developing and optimizing next-generation spintronic devices boasting unparalleled speed and efficiency.

Fundamental knowledge of the dynamics of magnetic domain walls (DWs) has been extensively studied [1-3] due to their potential applications for new spintronics-based information storage and processing devices [4-6]. Walker's study [7] based on the Landau-Lifshitz-Gilbert (LLG) equation has played a crucial role in our understanding of DW motion along magnetic nanowires, which indicates a well-known Walker solution for one-dimensional (1D) models [8,9]. However, this Walker solution can often not accurately describe experiments or numerical simulations on the field-driven DW motion, mainly when the DW structure varies. For example, inhomogeneous transverse DW that the width of DW varies in the transverse direction [10,11], and the curvature of DWs in perpendicularly magnetized layers [12,13] go beyond the rigid wall approximation of the 1D model. This discrepancy between theory and experiments attributes to Walker's ansatz that assumes a DW motion as a rigid body [2,7].

In this Letter, we propose a conceptually distinct method based on energy dissipation to study the DW dynamics accompanying resonance in an ultrathin film system with perpendicular magnetic anisotropy (PMA). The curved DW can be induced by the resonance of DW, expressed as a standing wave similar to its natural modes [14]. This resonance phenomenon is indirectly observed and is explained by numerical simulation and an analytical model using domain wall energy [12,13]. The resonance, which corresponds to distortion of the propagating DW profile, oscillates the DW in place; hence, it is indirectly involved in the DW velocity as energy dissipation and shall lead to a noticeable deviation of DW speed from the Walker model.

Micromagnetic simulations were performed to study the effect of the DW resonance on velocity. The simulation system consists of a thin film nanostrip whose geometry is $1000 \times 106 \times 0.6$ nm$^3$, the same as the one depicted in Fig. 1(a). For magnetic parameters of the given thin film, typical of Pt/Co/Pt system that exchange constant $A$ = 16 pJ/m, 1st uniaxial anisotropy constant $K_u$ = 1280 kA/m with an easy axis perpendicular to the interface, saturation magnetization $M_s$ = 1130 kA/m are adopted [15,16]. The exchange length $\sqrt{A/\mu_0 M_s^2} \approx 3.16$ nm; hence the mesh of 2 × 2 × 0.6 nm³ is used to take the finite difference method (FDM) as a numerical calculation method. The dynamic behavior of the magnetization structure is deduced by the solution of the LLG equation [17] for each mesh cell. Dzyaloshinskii-Moriya interaction (DMI) is ignored based on the symmetric structure of the Pt/Co/Pt layer system [18]. To manifest DW resonances, we selected damping parameter $\alpha$ = 0.01.

At the initial state, the left half of the strip is magnetized in the +z direction, and the right half is in –z. the Bloch wall magnetized in the +y direction is formed at the center of the strip. Here, a +z directional uniform DC magnetic field with a range from $H$ = 10 mT to 250 mT is applied on the strip to simulate the field-driven propagation of the DW. The magnetization structure moves slightly along the x direction to relocate the DW to the system's center at the end of each time step and minimize the effect of both ends of the strip [19,20]. The open-source micromagnetic simulation framework OOMMF [21] is chosen to implement the plan.

From the result of the simulation, one can represent the dynamic state of the DW as two profiles: $q(y,t)$, the position of the DW ($m_z = 0$), and $\phi(y,t)$, the azimuth angle of the magnetization vector over the $q$ profile. When the field is higher enough than a specific threshold called a Walker field, the time-averaged velocity of the DW in the PMA system is proportional to the strength of the perpendicular external magnetic field:

$$\overline{\dot{q}_{1D}} = \gamma \Delta \frac{\alpha}{1+\alpha^2} H, \tag{1}$$

where $\gamma = 1.76 \times 10^{11}$ rad·s⁻¹·T⁻¹ is the electron's gyromagnetic ratio, and $\Delta$ is the Bloch wall width. That equation is derived within the limits of the 1D model in that the length of the DW is short, but it also matches quite well on a long DW. However, figure 1(b) shows one example of the numerical simulation result that is out of that relationship. The solid green line represents the time dependence of the displacement of the DW under $H = 47$ mT, and the red dashed line represents that under 94 mT. The DW position is derived by the spatial average value of $q(y,t)$:

$$\langle q \rangle(t) = \frac{\int q(y,t)dy}{\int dy}. \tag{2}$$

Following the 1D approximation, the velocity of DW is proportional to the field strength; hence the velocity of DW under $H = 94$ mT should be twice faster as that under 47 mT. However, one can see the more rapid propagation of the DW under $H_z = 47$ mT than that under 94 mT. DW velocity $\overline{\langle \dot{q} \rangle}$ is deduced by the averaged time derivate of DW Position $\langle q \rangle$:

$$\overline{\langle \dot{q} \rangle} = \frac{\int \frac{d\langle q \rangle(t)}{dt} dt}{\int dt} \tag{3}$$

over a long enough time (> 20 ns) after the DW motion is stabilized since the DW moves back and forth under a high field. The inset graph shows the field strength dependence of the $\overline{\langle \dot{q} \rangle}$. Velocity under $H = 94$ mT (red circle) agrees well with the 1D approximation, but that under 47 mT (green square) exceeds. Figures 1(c) and 1(d) illustrate the magnetization structure and the time-variant DW profiles under $H = 47$ mT and 94 mT, respectively. Under 47 mT, the DW severely bends while propagating, but under 94 mT, it barely distorts.

Based on the above numerical simulations and reported DW dynamics experiments under a magnetic field, one can guess the effect of DW structure distortion on the dynamics of the DW. To verify that the variance of $q(y,t)$ and $\phi(y,t)$ are introduced:

$$\text{Var}(\tilde{q}) = \frac{\int (\tilde{q} - \langle \tilde{q} \rangle)^2 dy}{\int dy}, \tag{4a}$$

$$\text{Var}(\phi) = \frac{\int (\phi - \langle \phi \rangle)^2 dy}{\int dy}, \tag{4b}$$

where $\tilde{q} = q/\Delta$. These values can represent the degree of distortion of the DW position and azimuth angle, respectively. Figure 2(a) shows the variation of the DW under $H_z$ = 47 mT over a single period of motion (~ 0.76 ns). The variation of $\tilde{q}$ and $\phi$ arise and fall alternatively like orthogonal trigonometric functions. Magnetization structure images beneath Fig. 2(a) depict the varying shape of DW. When $\text{Var}(\tilde{q})$ is high, the q profile is bent, and when $\text{Var}(\phi)$ is high, the q profile is straight, but its $\phi$ profile is distorted.

After gathering the results over the entire field strength range, conditions for breaking the field-velocity relationship are revealed. Field dependence of $\overline{\langle \dot{q} \rangle}$ and $\overline{\dot{q}_{1D}}$ are illustrated in Fig. 2(b). Within the domain from $H$ = 10 mT to 250 mT, several velocity peaks that exceed $\overline{\dot{q}_{1D}}$ are found. The DW profile is heavily deformed and varies while propagating at the peaks, like the 47 mT case. However, as illustrated in the schematics above the graph, the distorted DW profile's shape differs for each velocity peak.

The relationship between DW resonance and its velocity can be interpreted by understanding the origin of DW displacement. Under the applied field, the DW moves toward where the domain aligned parallel to the field expands. In this process, the entire system loses its energy, to be exact, the Zeeman energy. In other words, energy pumped by the applied field into the system is dissipated by the dynamic motion of DW [22]. Significantly, the applied field, more robust enough than the Walker field of the DW, moves DW back and forth; that means the DW displacement mainly occurs by energy dissipation originated by the motion of DW.

One can quantify the energy dissipation of the DW by introducing generalized coordinates and taking the DW to the phase space [1,10-13]. The magnetization follows the open-end boundary condition [3], that is

$$\left.\frac{d\theta}{dy}\right|_{y=0,L} = 0, \left.\frac{d\phi}{dy}\right|_{y=0,L} = 0, \tag{5}$$

where $\theta$ and $\phi$ are the polar and azimuth angle of the magnetization vector, respectively, and

$L$ is the width of the strip. Hence $q(y,t)$ and $\phi(y,t)$ can be represented as the summation of cosine profiles [13,14]:

$$\tilde{q}(y,t) = \sum_{n=0} \tilde{q}_n(t) \cos\left(\frac{n\pi y}{L}\right), \quad (6a)$$

$$\phi(y,t) = \sum_{n=0} \phi_n(t) \cos\left(\frac{n\pi y}{L}\right), \quad (6b)$$

where $\tilde{q}_n \equiv q_n/\Delta$. Since the given system is ultrathin, the DW can be considered uniform in the z-direction. Here, the rate of change of $q_n$ and $\phi_n$ with $n \neq 0$ can be considered as that of DW shape, hence, the quantity total distortion variation rate that defined as

$$D \equiv \sum_{n=1} D_n = \sum_{n=1} \left(\overline{\dot{\tilde{q}}_n^2} + \overline{\dot{\phi}_n^2}\right), \quad (7)$$

which is the same as the velocity of the DW in 2n-dimensional phase space and can represent the amount of energy dissipation originated by DW resonance only.

Applying time domain Fourier transform to $\phi(y=0,t) - \phi_0(t)$ is valid to see the tendency of $D_n$ of resonance modes, since all kinds of cosine-shape open end standing wave consisting of the DW resonance form their antinodes on $y = 0$. Figure 3(a) shows the result of the Fourier transform over the field domain from 10 mT to 250 mT. Dark points mean intense resonance of mode proportional to its frequency. Although the $\phi_0(t)$ component subtracted from the $\phi(y=0,t)$ in the time domain, its trend remains in the form of a diagonal line following twice of $\overline{\dot{\phi}_0}$ due to its nonlinear nature. In the figure, all resonant points are aligned on a few horizontal lines where $f_n = n^2\omega_0/2\pi$ and $\omega_0/2\pi = 0.69$ GHz. Owing to the resonant wave interaction, forced mode with frequency $\dot{\phi}_0$ drives DW natural modes whose frequency $f_n$ is the same with an integer multiple of forced mode frequency, or the summation of their frequency is the same as that of the forced mode [23]. The given phenomenon originated from its nonlinear characteristic that can be verified from the asymmetric shape of velocity peaks. Figure 3(b) displays the field strength dependency of $D$ and $D_n$. The variation rate $D_n$ tends to form peaks with larger amplitude for high n values since the resonance frequency is proportional to $n^2$.

To understand $D_n$, one needs to reformulate the motion of DW using generalized $2n + 2$ coordinates $(q_0, q_1, q_2, \ldots, \phi_0, \phi_1, \phi_2, \ldots)$. The areal density of the DW Rayleigh dissipation function can be written as [24]

$$\mathcal{F} = \frac{\alpha M_s}{|\gamma|}\left(\frac{\dot{q}^2}{\Delta} + \Delta \dot{\phi}^2\right). \tag{8}$$

For simplification, $\Delta$ is regarded as constant. Using the definition of the generalized coordinates and the orthogonality of trigonometric functions, the areal density of the Rayleigh dissipation function can be reformulated as

$$\mathcal{F} = \frac{\alpha M_s \Delta}{2|\gamma|}\left(2\mathring{q}_n^2 + 2\dot{\phi}_0^2 + \sum_{n=1}\left((\mathring{q}_n^2) + \dot{\phi}_n^2\right)\right). \tag{9}$$

Now, suppose a DW moved sufficiently far distance during a long enough time. The energy loss caused by that displacement is approximately the same as the decreased amount of Zeeman energy, that is, $2M_s H \tilde{q}_0$ per unit area of DW. Neglecting the fluctuation of q, the average DW energy loss per unit of time and area is

$$\overline{\dot{\sigma}_{loss}} = 2M_s H \Delta \overline{\dot{\tilde{q}}_0}. \tag{10}$$

Meanwhile, the energy loss rate caused by damping is two times the time-averaged Rayleigh dissipation function,

$$\overline{\dot{\sigma}_{loss}} = 2\bar{\mathcal{F}} = \frac{\alpha M_s \Delta}{|\gamma|}\left(2\overline{\mathring{\tilde{q}}_n^2} + 2\overline{\dot{\phi}_n^2} + D\right). \tag{11}$$

hence, using the 1-D DW model equation [7]

$$\dot{\phi}_0 = |\gamma|H - \alpha \mathring{\tilde{q}}_0, \tag{12}$$

the equation (10) and (11) can be written as a quadratic equation about $\overline{\dot{\tilde{q}}_0}$:

$$(1+\alpha^2)\overline{\tilde{\tilde{q}}_0^2} - \frac{1+2\alpha^2}{\alpha}|\gamma|H\overline{\tilde{\tilde{q}}_0} + (|\gamma|H)^2 + \frac{D}{2} = 0. \tag{13}$$

Here, $D$ is identical to the stacked sum of the graph in Fig. 3(b), which represents the DW structure deformation rate. Note that it is identical to 1D model when $D = 0$. can be used to presume the velocity of DW, as shown in Fig. 3(c). The figure illustrates that the predicted velocity increment using equation (13) (dotted black line) is nearly close to the increment directly calculated from the simulation (solid orange line). That presumption does not quite match the peaks where $n = 4$ mode occurs, owing to the spin wave emission accompanied by $n = 4$ mode also arousing energy dissipation. Only the $n = 4$ mode emits spin wave since lower frequency spin wave cannot propagate through the domain. [25]

Figure 4(a) illustrates the stabilized trajectory of the DW in the $\tilde{q}_n - \phi_n$ phase space under $H_z = 47$ mT field. The DW draws a large limit cycle rotating in the CCW direction on the $\tilde{q}_2 - \phi_2$ phase space, but trajectories in the other $\tilde{q}_n - \phi_n$ spaces are significantly smaller. The frequency of that $n = 2$ limit cycle is 2.68 GHz $\approx 4\omega_0/2\pi$, which is twice the frequency of $\phi_0 = \langle\phi\rangle = \int \phi(y,t)dy/\int dy = 1.34$ GHz. That is due to the behavior of the DW as a parametric oscillator [12-14] driven by an inplane stray field proportional to $\sin 2\phi$. In this condition, the limit cycle is drawn as a circle. This trajectory of DW in the $\tilde{q}_n - \phi_n$ phase space is determined by mutual confrontation between dissipation and driven force, that is, stray field. In the phase space, distorted DW draws a stable spiral with angular frequency $n^2\omega_0$, falling to the origin with a radial velocity proportional to $\alpha$, but the resonance caused by stray field pushes the DW away from the origin (See supplement). The scale of the trajectory can be compared using its radius

$$r_n = \sqrt{\tilde{q}_n^2 - \phi_n^2}, \tag{14}$$

that depicted in Fig. 4(b). The $r_2$ value that is relatively larger than other $r_n$, and showing uniform tendency proves the manifestation and stabilization of the *n* = 2 mode. The schematic diagrams of the DW during the 2-period of the limit cycle are enumerated in Fig. 4(c). The DW alternatively changes its shape from the bent states ($|\tilde{q}_2| \gg 0$) to the twisted angle states ($|\phi_2| \gg 0$).

In conclusion, we have found that energy dissipation caused by the DW resonance is the origin of the DW velocity increment. The DW resonance can be treated as a parametric oscillator, oscillating in various shapes that can be decomposed into several standing waves. The amount of energy dissipation can be calculated from the variation rate of DW distortion, which is proportional to the velocity of DW in $\tilde{q}_n - \phi_n$ phase space. Quick motion in phase space means severe and frequent distortion of DW and hence, much energy dissipation. As a result, DW moves faster to converge the entire system to a minimum energy state where the DW reaches the end of the system. This study offers a method to interpret an oscillating magnetization by calculating its energy dissipation, which can be applied to other magnetic structures like magnetic skyrmion and magnetic vortex.

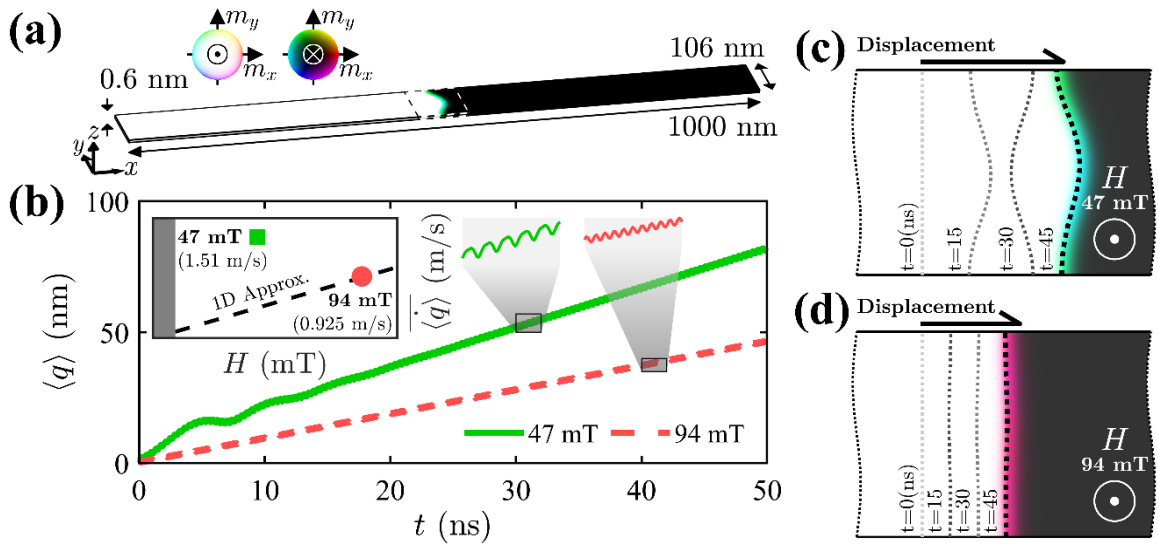

**FIG. 1** (Color Online). (a) The cobalt thin film nanowire schematic and its magnetization structure. The dashed line box indicates the area where the DW locates. (b) Time dependence of the DW displacement under magnetic field in +z direction. Green solid line $H$ = 47 mT, red dashed line $H$ = 94 mT. (inset) The field strength dependency of the time-averaged velocity of the DW. The black dashed line follows the 1D approximation of the velocity. (c), (d) The time-variance profile of the DW under $H$ = 47 mT and 94 mT, respectively. Each dashed line shows the line where $m_z$ = 0 at $t$ = 0, 15, 30, and 45 ns.

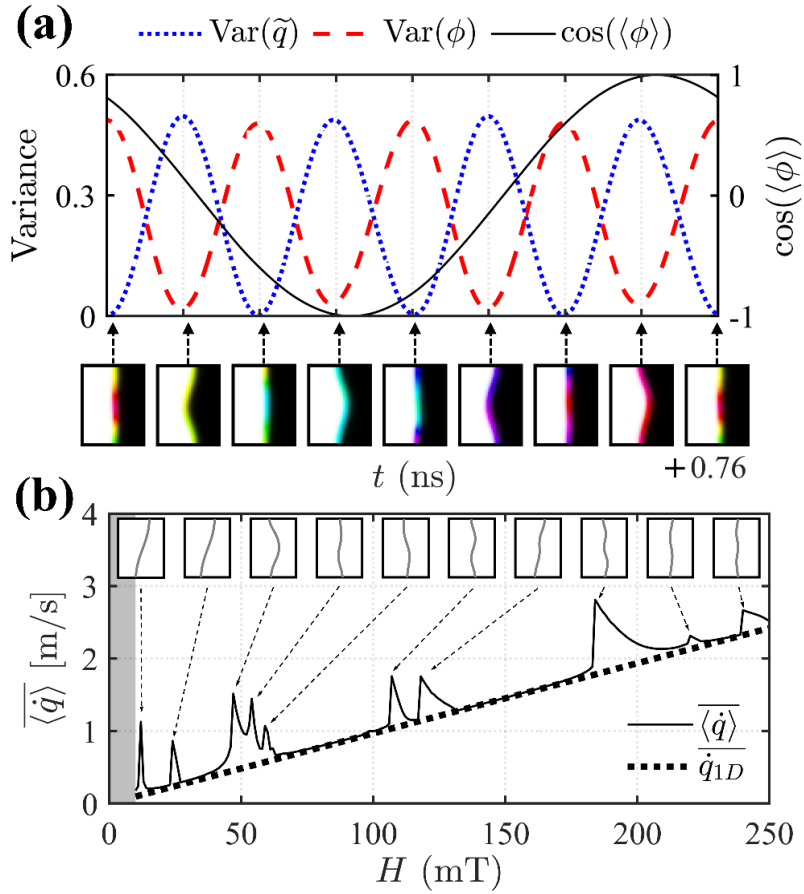

**FIG. 2** (Color Online). (a) The variation of the DW structure during a single period under $H = 47$ mT. The Blue dotted line follows the variance of the DW displacement, and the red dashed line follows the variance of the azimuth angle of the magnetization vector over the DW. The solid black line indicates the cosine of the average of the azimuth angles. Beneath images show the magnetization structure at each extremum of the variances. (b) Field strength dependence of the time-averaged velocity of the DW. The solid line is derived from the micromagnetic simulation, and the dashed line follows the 1D approximation of the velocity. The gray box on the left indicates the steady region, the transient region between the steady region and the precession region defined by Walker. (Inset) schematics of the most severely bent DW at each velocity peak.

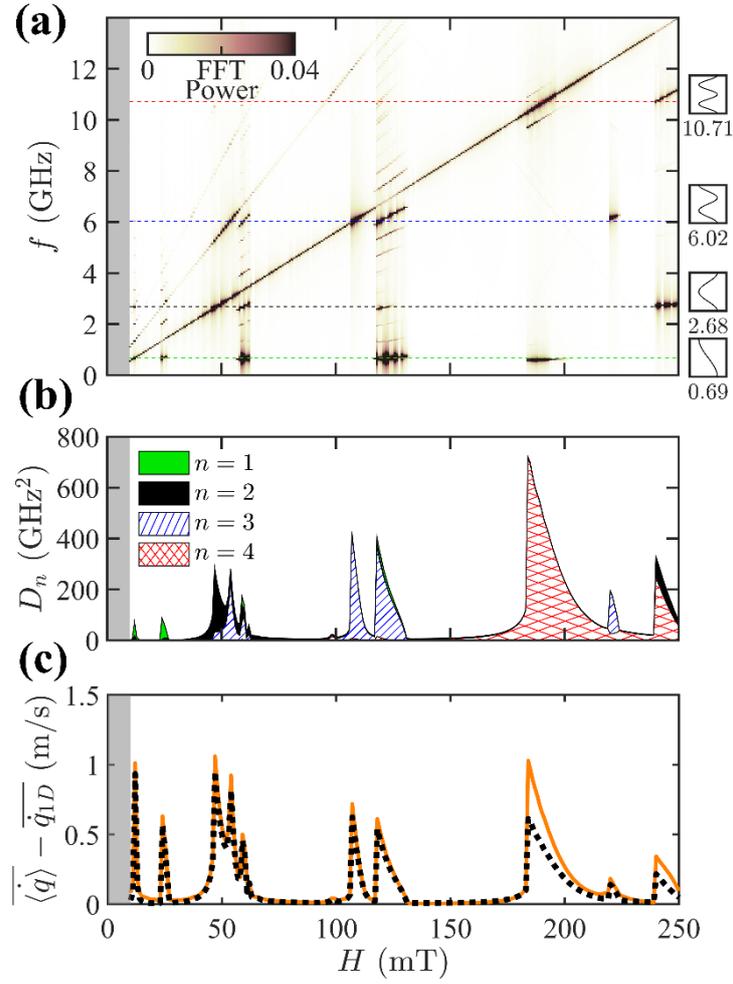

**FIG. 3** (Color Online). The spectrum of the DW and its relationship with velocity. (a) The field strength dependence of the time domain FFT strength of $\phi(y=0,t) - \phi_0(t)$. Green, black, blue, and red dotted lines indicate the frequency of the DW normal modes with $n$ = 1, 2, 3, and 4, respectively. (b) Stacked graph of the time-averaged DW distortion variation rate $D_n$. The height of the graph depicts the total DW distortion rate $D = \Sigma D_n$. (c) The difference between the 1D approximation of the time-averaged DW velocity and the micromagnetic simulation results. DW velocity is derived from the simulation in two ways: estimation using the total DW distortion rate (Dotted black line) and direct calculation from the magnetization structure (solid orange line).

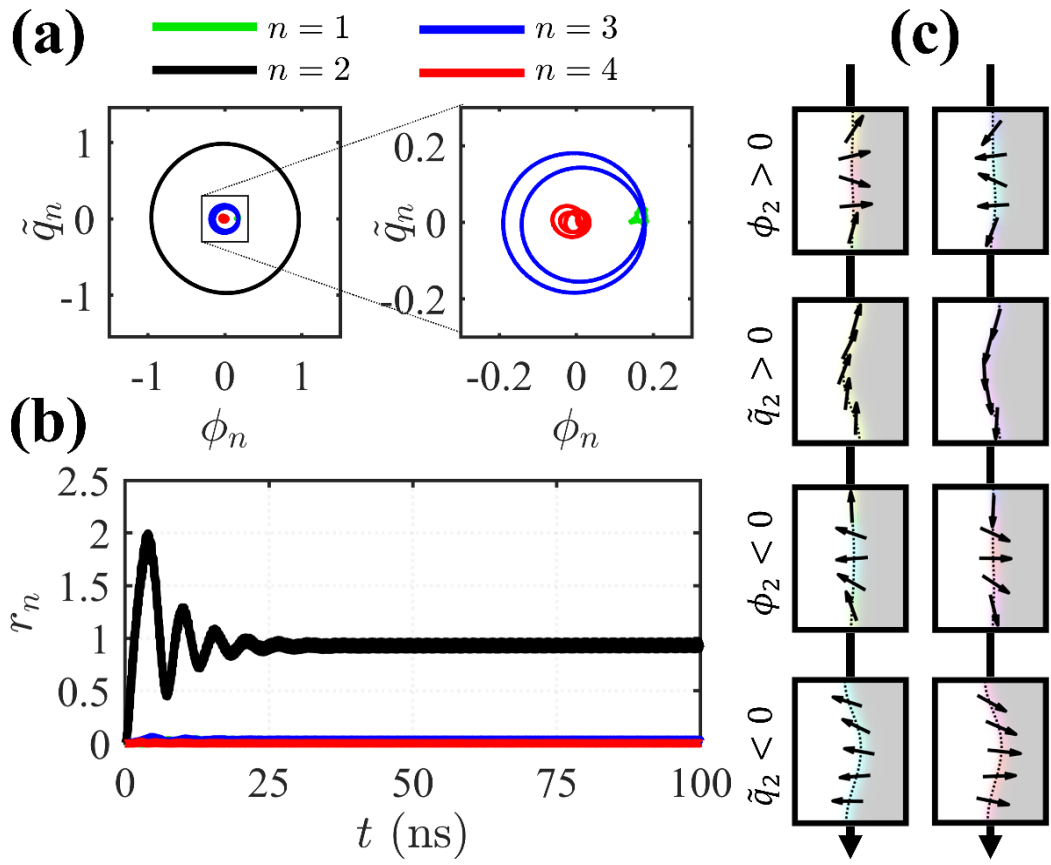

**FIG. 4** (Color Online). The dynamics of the DW under $H$ = 47 mT inside the $\tilde{q}_n - \phi_n$ phase space. (a) The limit cycle of the DW inside the phase space. (b) Time dependence of the strength of the DW distortion $r_n = \sqrt{\tilde{q}_n^2 + \phi_n^2}$. (c) Snapshots of magnetization during two periods of motion inside the phase space. Each vertical line shows two periods that draw a single $n$ = 2 limit cycle and emerge alternatively.

* ganghwi@unist.ac.kr

† kisuk@unist.ac.kr